\documentclass[12pt]{iopart}
\usepackage{graphicx}
\usepackage{amsmath}
\usepackage{slashed}
\usepackage[normalem]{ulem}
\usepackage{hyperref}
\usepackage{color}
\definecolor{gold}{rgb}{0.83, 0.69, 0.22}
\definecolor{green}{rgb}{0.0,0.5, 0.0}

\newcommand{\bF}{\textbf F}

\begin{document}

\title{Training models using forces computed by stochastic electronic structure methods}

\author{David M. Ceperley, Scott Jensen}
\address{Department of Physics, University of Illinois, Urbana, Illinois 61801, USA}
\ead{ceperley@illinois.edu}

\author{Yubo Yang}
\address{Center for Computational Quantum Physics, Flatiron Institute, New York 10010, NY}
\author{Hongwei Niu}\address{Ningde Amperex Techology Ltd, Ningde, Fujian 352100, China}
\author{Carlo Pierleoni}\address{Department of Physical and Chemical Sciences, University of L'Aquila, Via Vetoio 10, I-67010 L'Aquila, Italy}
\author{Markus Holzmann}\address{Univ. Grenoble Alpes, CNRS, LPMMC, 38000 Grenoble, France}

\vspace{10pt}

\begin{abstract}
Quantum Monte Carlo (QMC) can play a very important role in generating accurate data needed for constructing potential energy surfaces. We argue that QMC has advantages in terms of a smaller systematic bias and an ability to cover phase space more completely. The stochastic noise can ease the training of the machine learning model. We discuss how stochastic errors affect the generation of effective models by analyzing the errors within a linear least squares procedure, finding that there is an advantage to having many relatively imprecise data points for constructing models. We then analyze the effect of noise on a model of many-body silicon finding that noise in some situations improves the resulting model. We then study the effect of QMC noise on two machine learning models of  dense hydrogen used in a recent study of its phase diagram. The noise enable us to estimate the errors in the model.  We conclude with a discussion of future research problems.
\end{abstract}

\vspace{2pc}
\noindent{\it Keywords}: quantum Monte Carlo, potential energy surface, machine learning

%
%

\section{Introduction}
A key technology in the today's materials science and related fields is the ability to perform simulations of atoms and molecules using Molecular Dynamics (MD) or Monte Carlo (MC) methods.
Because electrons are more than a thousand times lighter than the nuclei, at room temperatures we can often assume that electrons are in their ground state. We can then ``integrate them out'' and deal with the motion of the nuclei alone by using the Born-Oppenheimer separation between electronic and nuclear degrees of freedom. 
Once the electrons have been eliminated, the resulting ``potential energy surface'' (PES), $V(R)$, is a function of 3$n_I$  coordinates of the $n_I$ nuclei.
Traditionally this potential has been approximated by an empirical function, such as a pair potential, or calculated `on-the-fly' using density functional theory~\cite{Car1985}.

Much progress has been made in approximating the PES using a neural network or machine learning (ML) model by Behler~\cite{Behler2007}, Bartok~\cite{Bartok2010}, Zhang~\cite{Zhang2018,Zhang2018a, Han2018, Jia2020}, Chmiela~\cite{Chmiela2017} and many others. 
In this approach, the ML model is trained (or fit) to a set of energies and forces of typical configurations of the modeled system. The data often comes from a density function theory (DFT) or other electronic structure method. However, such data can be biased by approximations used in those methods. 

We examine in this paper whether stochastic methods, such as variational and diffusion quantum Monte Carlo can be used to provide the training data so that the resulting model will be more accurate.
Such an approach has been considered previously, for example in work by Alfe~\cite{Alfe2013}, Sorella~\cite{Tirelli2022, Tenti2023},  Niu~\cite{Niu2023}, Huang~\cite{Huang2023}, Nakano~\cite{Nakano2022} and Cheng~\cite{Cheng2020}. But several aspects of this approach, such as the effects of the noise on the model, have not been discussed, to our knowledge.

In the following we introduce QMC methods, and discuss various errors that enter into the models and their relationships.  We then discuss several examples such as those based on linear regression, a ML model for silicon and for dense hydrogen. We conclude with observations on future research questions. 

\section{Quantum Monte Carlo Methods for electronic structure}
We briefly review some QMC algorithms that estimate electronic properties using a stochastic process. We write $R=\{s_1,s_2 \ldots s_{n_I} \}$ for the $n_I$ nuclear coordinates with charge $Z_j$ on the $j^{th}$ ion  and $r=\{r_1,r_2 \ldots r_{n_e} \}$ for the $n_e$ electron coordinates. The non-relativistic electronic Hamiltonian in atomic units is:
\begin{equation} \label{eq:ham}
\hat{H} = \sum\limits_{i=1}^{n_e} \left[
-\frac{1}{2}\hat{\nabla}^2_i
-\sum_{j=1}^{n_I} \frac{Z_j}{| r_i-s_j |}
\right]
+\sum_{i<j} \frac{1}{| r_i-r_j|}.
\end{equation}
 $R_{\alpha}$ will refer to the $\alpha^{th}$ sampled configuration (a $3n_I$ vector) used for training or evaluation. 
 By $\bF_j (R)=-\nabla_j V(R)$ we mean the $3$ vector of forces on the $j^{th}$ ion for a configuration R.
 
 We mention the QMC methods that we think will be useful for generating machine learning data. For a detailed discussion of the various methods see\cite{Martin2016}.
\begin{description}
    
    \item [Variational Monte Carlo (VMC)] 
    
Metropolis (i.e. Markov Chain MC or MCMC)~\cite{Ceperley1977} is used to sample electronic coordinates from the square of a trial wavefunction $|\Psi_T(r|R))|^2$  
and calculate its (variational) energy:
    \begin{equation}
    E_V(R) =\frac{\int dr \Psi_T^*(r|R) {\hat H} \Psi_T (r|R) }{\int dr |\Psi_T (r|R)|^2 }.
    \end{equation}

The simplest correlated trial function used in QMC is the Slater-Jastrow form:
    \begin{equation}
        \Psi_T (r|R) = \exp(-U(r|R) ) \det_{ik} (\phi_k (r_i|R))
    \end{equation}
where $U(r|R)$ is a bosonic Jastrow factor, and $\phi_k (r|R)$ is a set of single-electron orbitals. Both of these terms are varied to minimize $E_V(R)$.
There has been recent work on machine learning trial wave functions \cite{Ruggeri2018,Xie2022,Pescia2022,Li2022,Wilson2023,Xie2023} which achieve excellent variational bounds.  However, the current methods are quite time consuming, requiring very long optimizations. We argue that ML needs  ``lightweight'' QMC methods in order to provide sufficient training data. For example, in dense hydrogen we work with analytic forms requiring little optimization of trial functions~\cite{Pierleoni2008}. 
Forces are then calculated as the derivative with respect to the ionic coordinates, invoking the Hellman-Feynman theorem~\cite{Feynman1939}.
    
\item [Projector MC] 
     takes a trial wavefunction and uses the Hamiltonian to project out errors of the trial function: 
    \begin{equation}
        \Phi=\lim_{t \rightarrow \infty } e^{-t {\hat H}} \Psi_T.
    \end{equation} 
There are several methods by which this can be accomplished. Since electrons are fermions, the methods must employ an ansatz to control the stochastic error for many-electron systems 
such as the fixed-node~\cite{Ceperley1980} or fixed-phase~\cite{Ortiz1993}  approximations. Fixed-node and fixed-phase methods solve exactly for the modulus of the wavefunction within statistical errors but the bias from the ansatz can be significant~\cite{Martin2016} if only a single Slater determinant is used for antisymmetry. 
One can use signful methods~\cite{ceperley1984} to correct this error.
    
    {\bf Diffusion MC (DMC)}\cite{Ceperley1980} accomplishes the projection by representing the system as an ensemble of many-body configurations that execute a drifting, branching random walk. DMC has a mixed estimator bias for forces; there are several procedures for correcting this bias.
    
    {\bf Reptation Quantum Monte Carlo (RQMC)} holds a segment of a many-body path in the computer memory and adds or subtracts from the path using a modified MCMC \cite{Baroni1999,Pierleoni2005}.
If the projection time is large enough, forces taken in the middle of the path do not have the mixed estimator bias but they will still have the fixed-node error. 
    
    \item [Path Integral] methods~\cite{Ceperley1996} work at non-zero temperature and differ from RQMC in working with closed paths in configuration space.
     PIMC methods could address important problems for warm dense matter addressing a regime that is difficult for most electronic structure methods. However, PIMC has only been made practical for temperatures above some fraction ($25\%$) of the Fermi temperature.  Sampling is done  with MCMC and must include sampling electron permutations. Early work calculated forces between protons in an electron bath~\cite{zong1998}.

\end{description}

For ML the ability to compute forces is crucial. There are several methods for computing forces with QMC\cite{Assaraf2000,Chiesa2006}.     
For generating ML data the efficiency, namely how much computer resources are needed to reach a given error on the forces is important. More research is needed on the relative efficiencies, the biases, and the appropriate domains of the QMC methods for the calculation of accurate forces. 

In the past, many researchers concluded that the computer resource requirements to compute forces using QMC were excessive.
However, available resources continue to grow year-by-year while the cost of human resources needed to come up with reliable predictions is not decreasing. A more reliable electronic structure method can reduce this human cost and thereby allow for more efficient simulations of materials.
One of the main roadblocks in the past has been the availability of high quality QMC codes but this situation has improved with development of such packages  as QMCPACK\cite{Kim2018,Kent2020}, TURBORVB\cite{Nakano2020}, and CASINO\cite{Needs2020}. In addition to more plentiful computer resources, QMC techniques are improving. We believe their efficiency could be increased much further as we discuss below, especially for the problem of determining unbiased forces.

QMC, in general, is more accurate than other electronic structure methods.  By that we mean that it has a lower bias than other methods such as those based on DFT where the choice of the functional is guided by external experimental knowledge. Experimental data of high quality is rarely available in many applications. 
Its low bias is the most important reason to use QMC to generate data. 
QMC is also a more general technique since it is able to handle highly correlated systems such as lattice models, superconductors, quantum crystals, and liquid helium.

The continuum basis set  used in VMC, DMC, RQMC and PIMC is better suited for disordered systems, such as a liquid, since distortions caused by a finite basis set are absent. For deterministic methods, one needs to assure convergence in the basis set, e.g. the plane wave cutoff or the expansion in local basis functions. 

A very important advantage of stochastic methods is that extra degrees of freedom that need to be integrated over do not necessarily increase the computational effort.  For example, twist averaging~\cite{Lin2001}, is used to eliminate some finite size effects by integrating over the Brillouin zone of the super-cell. If we compute QMC properties at $P$ twist values, this does not increase the QMC computation effort since the $P$ points all serve to reduce the statistical error. Instead, within deterministic methods such as DFT, this will increase the computer time by a factor of $P$. Another example of this advantage occurs in the path integral algorithm for quantum protons which requires the electronic energy of several replica of the protons to accept an update~\cite{Pierleoni2008}.
The same logic applies to getting data to train a ML model. 
The forces at a large set of configurations can be obtained using the same computer budget  as a small set of configurations although the stochastic errors of the large set will be larger. But as we show below, the large set is better for constructing the ML model.  Hence there is less importance to selecting an optimal set of configurations to use in training the model. 

We expect that important applications of machine learning will be to complex, multi-component, many-body systems, for example a ternary liquid.  
For those systems, one does not know how many configurations will be required, how to place them,  and how to assure that the energies are unbiased.    
With QMC, one can oversample the configuration space.
Even if many of the points are clustered, the computational effort serves to reduce the error in the fitted potential energy surface. 
We do not wish to imply that methods to place the QMC points in an efficient manner will not improve the efficiency but that optimal placement is not always necessary. 

Noisy data is also good for training neural networks and is often added to deterministic methods to aid in optimization.  It can help to find the best model more easily~\cite{Hao_Visualizing_2018}.
With noisy data, one does not need to worry about data clustering: one will not have a problem with oversampling.  

The QMC computer time scales well with the number of electrons compared to other methods such as coupled cluster. For a recent discussion of the scaling and accuracy of various electronic structure methods see ref.~\cite{Motta2017}.  
For systems with many hundreds of delocalized electrons the computation of determinants to enforce antisymmetry will cost $n_e^3$ operations to do a multiparticle move. This is the same scaling as in DFT algorithms.  (Note than one is calculating $3n_I$ forces during this move.)  
In practice, because computing the determinant is a simple linear algebra operation, QMC time is typically dominated by other operations such as calculation of the charge-charge interaction and scales more slowly than $n_e^3$.  
QMC can easily profit from massively parallel computer environments since memory demands can be kept modest.

We will examine some of these issues in the rest of the article. We focus on methods that do not have basis set errors (i.e. work in the continuum) and work in the electronic ground state, that is the VMC, DMC and RQMC methods.  

\section{ Errors in Quantum Monte Carlo and Machine Learning Models}
Let $V(R)$ be the exact energy (i.e. the ``ground truth'')  and $\bF_k (R)=-\nabla_k V (R) $ the corresponding force on the $k^{th}$ atom.
The energy and forces will be estimated using a method for numerically solving the ground state Schr\"odinger equation for the nuclei in a given position obtaining for a set of $M$ configurations $\{ R_{\alpha} \}$ ($1 \leq \alpha \leq M$) with estimated values  for the potential (${\hat V_{\alpha}}$)  and  forces (${\hat \bF_{k}}$). Using this data we then make a model $V_m(R)$. 
To help with further discussion we explicitly list and discuss the errors and their relations. Although the listed formulas below are for energies, the analogous errors in the forces are more important in ML. 

\begin{figure}[ht]
\begin{minipage}{0.4\textwidth}
\includegraphics[width=0.95\linewidth]{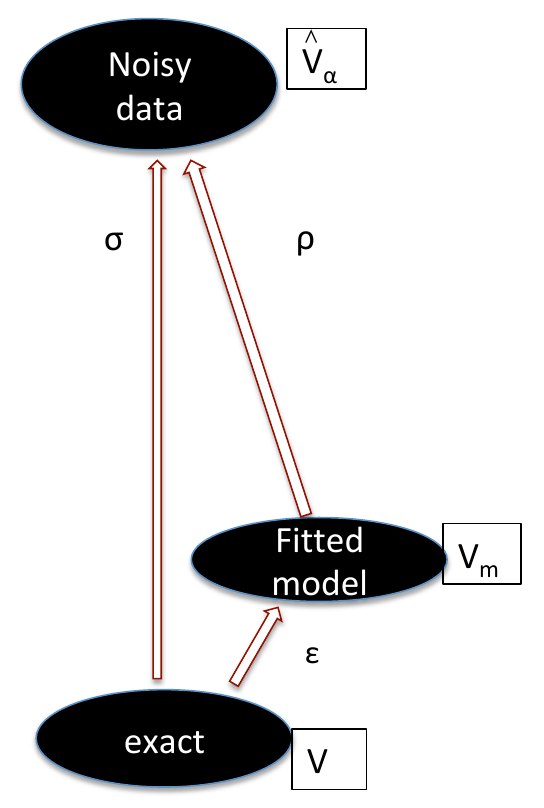}
\end{minipage}
\begin{minipage}{0.65\textwidth}

\begin{align}
{\vec \sigma}_V \equiv & {\hat V_{\alpha}}- V(R_{\alpha})\\
{\vec \rho}_V \equiv & {\hat V_{\alpha} }- V_m(R_{\alpha})\\
{\vec \epsilon}_V \equiv & V_m(R_{\alpha})- V(R_{\alpha})={\vec \sigma}_V-{\vec \rho}_V\\
\sigma_V= &\sqrt{\frac{1}{M}\sum_{\alpha} ({\hat V}_{\alpha}- V(R_{\alpha}))^2}\\
     \rho_V= &\sqrt{ \frac{1}{M}\sum_{\alpha}({\hat V}_{\alpha}- V_m(R_{\alpha}))^2 } \label{eq:rho_V}\\
     \epsilon_V =& \sqrt{ \frac{1}{M}\sum_{\alpha}(V_m(R_{\alpha})- V(R_{\alpha}) )^2 } \label{eq:eps_scalar}\\
\chi^2_V = & M \left[\frac{\rho_V }{\sigma_V}\right]^2  \qquad
\chi^2_F = 3 n_I M \left[\frac{\rho_F }{\sigma_F}\right]^2 
\end{align}

\end{minipage}
\caption{The left diagram shows the error notation used in this paper, with the defining equations on the right. The errors $\{ \vec{ \sigma}_V,{\vec \rho}_V,{\vec \epsilon}_V \}$  are vectors whose components are defined on each configuration $R_{\alpha}$. Their normalized moduli ($\sigma_V,\rho_V$,$\epsilon_V$)  are defined as the root mean square of the corresponding vector (i.e. divided by the number of configurations). For the analogous force errors, ($\vec{\sigma}_F,\vec{\rho}_F$,$\vec{\epsilon}_F$), one sums over the $3n_I$ Cartesian components of the mean squared differences as well as over the configurations and divides by $3n_IM$. }
\label{fig:Diagram1}
\end{figure}

\begin{description}
\item [The model error] is the deviation of the model with respect to the ground truth: $\epsilon(R)=V_m(R)-V(R)$.
Even with good data and training, unless the model is a sufficiently general function of $3n_I$ variables, the  model error will be non-zero.
 
\item [The stochastic error] is the difference between the QMC value (with noise) and the exact value:
$\sigma_{\alpha}={\hat V}_{\alpha}-V(R_{\alpha})$.
For simplicity, we assume that QMC is unbiased so that $\langle \sigma_{\alpha} \rangle = 0$: after enough averaging one will recover the exact energies and forces.  
Estimates of the stochastic errors on energies and forces are determined from the fluctuations in the QMC run, from the same data as the mean values and will depend on the ionic coordinates. An example of QMC noise for hydrogen is shown in Fig. (\ref{fig:Hydrogen_errors}).

\item[Fitting error] tells us whether the model fits the noisy, but unbiased data:
$\rho_{\alpha}={\hat V_\alpha}-V_m(R_{\alpha})$. 

\item[Chisquare] is the negative logarithm of the likelihood that the model fits the QMC data:  $\chi^2=\rho^2/\sigma^2$.

\item [Selection error.]  Since the assumed model is not perfect, it matters how many configurations $M$ are selected for training and where they are placed. We will assume that they are selected independently from a probability density $F(R)$ by using a stochastic process such as MC or MD.  
How exactly they should be selected (e.g. the function $F(R)$) is an important question which we will not address in this paper. Note that probability distribution for assessing the fitting and model errors, the testing distribution, could be different from $F(R)$. 

\item[Inter-model] or cross-mode validation error] is the difference in the energies and forces between two fitted models ($m1$ and $m2)$: ${ \delta} (R)={ V_{m1}} (R) -{ V_{m2}}(R) $.  It gives an indication of the importance of the stochastic error, the selection error, the model error and the fitting robustness on the models. 
 
\item [Bias of the data] means that $\left< \sigma_{\alpha} \right> \neq 0$.  This could either be a result from controlled errors, those that can be studied and eliminated such as time step error, basis set error, convergence errors, or uncontrolled errors such as the fixed-node error of projection quantum Monte Carlo. 

\item [Finite-size errors] result from data being generated for small systems but used in simulations for larger systems.  The QMC calculations are on systems containing hundreds to thousands of electrons so that corrections are needed \cite{Lin2001,Chiesa2006,Holzmann2016} to use the data of models for larger systems.
\end{description}

One of the goals of Uncertainty Quantification is to estimate how accurate the model is. We can easily estimate $\sigma$ and $\rho$ but $\epsilon$ is more difficult since we do not have precise estimates of the energy or force for many-body systems. 
We can consider these 3 errors to be vectors with $N_c$ components where $N_c=M$ for the energy errors and  $N_c=3n_IM$ for the force errors. In this paper we define the norm of an error vector as 
$\sigma=\sqrt{\frac{1}{N_c}\sum_{\alpha} \sigma_{\alpha}^2}$.
Thus we define \{$\sigma_V$, $\epsilon_V$, $\rho_V$, $\sigma_F$, $\epsilon_F$, $\rho_F$ \} as detailed in Fig. (\ref{fig:Diagram1}).
If we define the vectors on a common set of configurations, the energy and force norms will obey the triangle inequalities:
\begin{equation}
  |\rho-\sigma| \leq \epsilon \leq \rho+\sigma.
\end{equation}
The norms should converge to a limit as $M \rightarrow \infty $ so that the inequalities will apply in that limit. In that limit we expect that $\epsilon_V$ will converge to
\begin{equation}\label{eq:epsexact}
    \lim_{M \rightarrow \infty} \epsilon_V = \sqrt{\int dR F(R) ( V_m(R)-V(R) )^2}.
\end{equation}
A finite number of configurations is a stochastic (i.e. Monte Carlo) evaluation of this integral.
We can construct other triangle inequalities using different models.

We can make tighter bounds on the model error, $\epsilon$, by considering how the fluctuations in the QMC data could be correlated with the errors of the model.  
Suppose we assume ${\vec \epsilon} = \eta{\vec \sigma}+{\vec \epsilon}_0$ where ${\vec \epsilon}_0$ is the model error vector with zero noise $\sigma=0$. In the most extreme procedure the model could consist of taking the closest configuration so that $\eta=1$. For a good model which is relatively unaffected by the QMC noise because the noise is filtered or averaged over, and for models such as those based on a DFT functional or an empirical potential, we expect $\eta \approx 0$.  
Because the vector of QMC errors, ${\vec \sigma}$, is uncorrelated with the ideal model errors ${\vec \epsilon}_0$:
\begin{equation}\label{eq:rhoepsneta}
    \rho^2 = \epsilon^2+\sigma^2  -\frac{2}{M} \sum_{\alpha} \sigma_{\alpha} \epsilon_{\alpha}=  \epsilon^2+ (1-2\eta)\sigma^2 \pm 2 \sigma \epsilon_0 /\sqrt{N_c}.
\end{equation}
In the case of plentiful data, we expect $\eta \rightarrow 0$, so that $\epsilon \rightarrow \sqrt{\rho^2-\sigma^2} \pm \sigma /\sqrt{N_c}$.
In general, we expect that $0\leq \eta \leq 1$,  leading to tighter bounds on $\epsilon$ in terms of $\sigma$ and $\rho$:
\begin{eqnarray}
    \sqrt{\rho^2-\sigma^2} &\leq \epsilon \leq \sqrt{ \rho^2+\sigma^2} &\qquad \sigma \leq \rho \nonumber \\
    |\rho-\sigma| &\leq \epsilon \leq \sqrt{ \rho^2+\sigma^2} & \qquad \sigma \geq \rho.
\end{eqnarray}

To estimate the statistical uncertainties within a model, we can use a K-fold cross-validation, a variation of the jackknife procedure in statistics. One partitions the set of configurations into K ``folds'' possibly stratifying  the selection of the folds so that each fold contains roughly the same number of configurations at a given density and temperature. The number of folds may be limited by computer resources since each fold requires a model optimization. 
From the folds we make K training sets by sequentially dropping a fold from the complete data set. Using those training sets one finds K models $V_{mk}(R)$ and an average model $V_{m0}(R) = \sum_k V_{mk} (R)/K $.
Summed over a test set of configurations, one finds the rms differences in the energies and forces between the models and the average model. The jackknife estimated model error for the energy is:
\begin{equation} \label{eq:delta_V}
\delta_V=\sqrt{\frac{K-1}{KM} \sum_{k,\alpha}^{K,M} (V_{mk}(R_{\alpha})-V_{m0} (R_{\alpha}))^2
}
\end{equation}
with a similar relation for $\delta_F$ but normalized by an additional factor $3n_I$.  These errors estimate the overall effect of the stochastic QMC errors, the selection errors and optimization errors, i.e. how would the energies and forces change if one redid the complete analysis with M newly sampled configurations, new QMC calculations on those configurations and a new optimization.  They do not include model errors.

The stochastic and selection errors come from the finite sampling and can be reduced by more sampling. 
The model error can be estimated by showing that the deviation of the model from the data is larger than expected on the basis of Eq. (\ref{eq:rhoepsneta}). 
To quantify the bias and finite size errors, one needs to perform systematic studies of convergence, for example by changing the trial wavefunction or varying the number of atoms. 

\section{Least Squares Analysis}
The simplest procedure for finding a model to represent the QMC data is to perform a linear least squares fit (LLSF), i.e. a regression. Some leading ML models such as Gaussian Approximation Potentials (GAP) ~\cite{Deringer2021} are based on LLSF.  Non-linear models can be approximated by linear models near their solution.  LLSF is a standard numerical method  
with a guaranteed single solution and methods for assessing the resulting error bars on the model.

\begin{figure}[ht]
\begin{minipage}{0.75\textwidth}
\includegraphics[width=0.95\linewidth]{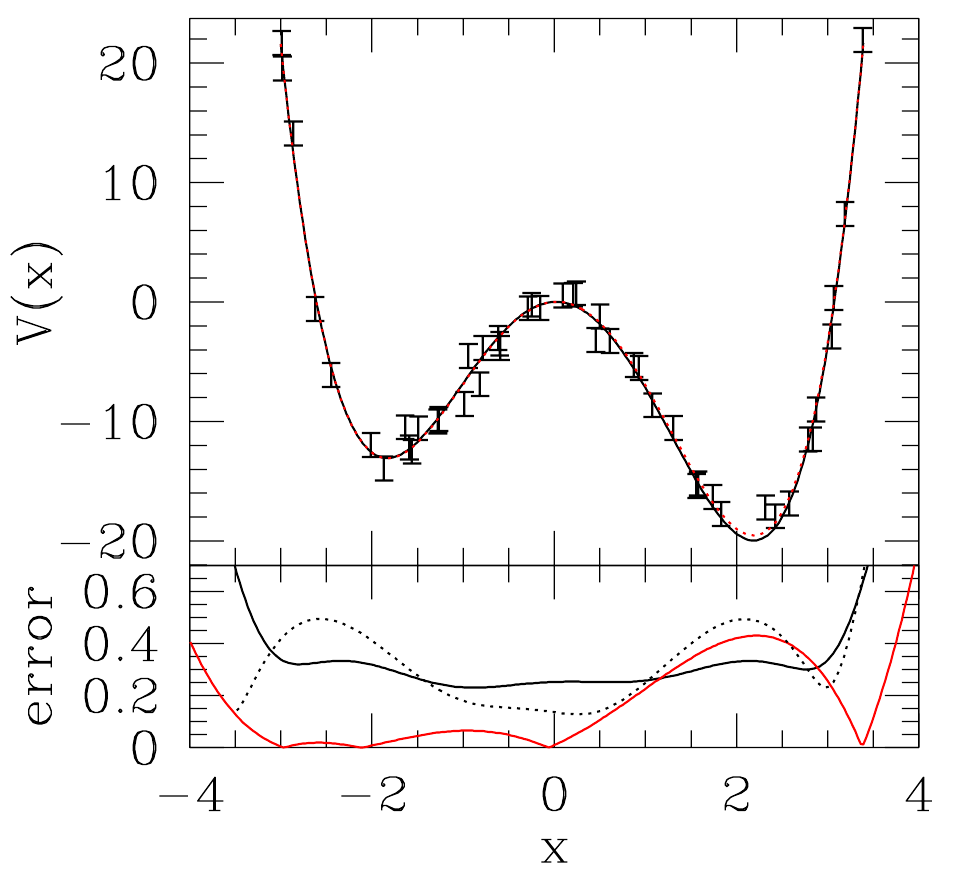}
\end{minipage}
\begin{minipage}{0.2\textwidth}
    \begin{tabular}{|c|c|}
    \hline
        n & $\chi^2$ \\
        \hline
         2&6216  \\
         3&5333  \\
         4&44  \\
         5&44 \\
         6&43 \\
         7&42 \\
         8&41 \\
         \hline
    \end{tabular}
\end{minipage}
\caption{Left top panel: Exact potential energy $V(x)$ (black curve) and estimated model $V_m (x)$ (red dots).  The data with error bars are the points used in the fit. Left lower panel:  model error $\epsilon(x) = V_m(x)-V(x)$ (red) and expected statistical error (black) from  eq.(\ref{eq:llsq}). The estimated error from 5-fold cross-validation is the dotted curve. 
Right table: $\chi^2$ versus polynomial degree, $n$.  Fits are satisfactory when $\chi^2 \approx 50-n  \pm 10$ since there are 50 data points. Fits for $n \ge 4$ are good.   Values of $n > 4$ have overfit the data since they have extra parameters without decreasing $\chi^2$. }
\label{fig:npoly}
\end{figure}

We illustrate how having stochastic error affects the results with a simple 1D quartic function,  shown in figure (\ref{fig:npoly}), by using LLSF with a polynomial basis.  
We assume that the parameters of the quartic potential (the ground truth) are unknown.
The 50 points with error bars represent QMC results for locations chosen at random and uniformly in the range $-3.5 < x < 3.5$.  To the ``exact'' but unknown potential, we added normally distributed noise with a standard deviation of $\sigma_V=1$. 
The noisy potential energies were then fit to a polynomial of degree $n$ using LLSF. The fit is the red dotted curve for a quartic polynomial, i.e. $n=4$. 
The QMC derived potential energy surface is more accurate than the individual QMC energies and is only different from the exact PES around $x=2$ by a maximum of 0.4 even though the data had a standard error of 1.  

The average error defined in Eq. (\ref{eq:epsexact}), computed by propagating errors in the data to errors in the solution (see Eq. (\ref{eq:llsq})) is found to be $\epsilon_V = 0.326$. Using 5-fold cross-validation we found $\epsilon_V = 0.346$. By fitting to a quartic polynomial, we have assumed that the PES must be smooth. If we had fit to a higher degree polynomial as shown in table ~(\ref{fig:npoly}), the errors would not have decreased.  
Using the value of $\rho,\sigma,\epsilon$ we can compute the sensitivity parameter and find $\eta=0.11$. 
This is reasonable since we have 5 free parameters in the fit with 50 data points so $\eta \approx 5/50=0.10$. 

For LLSF we want the amount of data to be greater than the number of free parameters.
To determine a quartic polynomial, exact energies at five distinct points would have been enough.  A deterministic method on this particular low-dimensional problem would, of course, be preferred for generating data since one can saturate the configuration space with accurate noise-free data.  However, one cannot saturate the space for realistic models of bulk materials.   

The table in Fig.~\ref{fig:npoly} shows $\chi_V^2$ vs polynomial degree. One has extremely poor fits using quadratic or cubic polynomials, but using a quartic polynomial is excellent. Increasing the order beyond this does not improve the fit and is overfitting the data.  The computed error, shown in the bottom graph, is excellent in the region $-3 < x < 3$  but it degrades outside of the region where data is present. 

Now consider a general least squares fit (LLSF) in a space of arbitrary dimension.   
Expand the model as $ V_m(R) = \sum_1^P a_k f_k (R)$ where  $f_k(R)$ are $P$ basis functions.  
Suppose we have unbiased energy data consisting of $M$ configurations $\{R_{\alpha}\} $ sampled from a distribution $F(R)$ with estimated potential  energies $\hat{V}_{\alpha}$  and standard errors $ \sigma_{\alpha} $. 
We  determine $a_k$  by minimizing the likelihood $\chi^2=\sum_{\alpha=1}^M ((V_m (R_{\alpha})-\hat{V}_{\alpha})/\sigma_{\alpha})^2$  with respect to $a_k$.  
The solution is determined by solving linear algebraic equations in the $P$ unknowns ~\cite{NumRecipes}. 

The variance of the energy of the fitted model at a point $R$
\begin{equation}
    \delta_V(R)^2 = \left<  (V_m(R)-\left< V_m(R) \right>)^2\right> =\sum_{mn}f_m(R) [{\hat A}^{-1}]_{mn}f_n(R)
    \label{eq:llsq}
\end{equation}
is given in terms of the inverse of the matrix:
\begin{equation}
    {\hat A}_{mn} = \sum_{\alpha=1}^M \frac{f_m(R_{\alpha})f_n(R_{\alpha})}{\sigma_{\alpha}^2}.
\end{equation}
The above brackets $\left< \ldots \right>$ mean that one averages over both the electronic structure process that determines the noise $\sigma_{\alpha}$ and the selection of points $R_{\alpha}$ sampled from $F(R)$.

Let us examine how $\delta (R)$ depends on the ``running'' computer time $T$ (i.e. the computer time when you are actually generating forces and energies. We exclude from $T$ the start-up time for the QMC process). We assume that the variance of the QMC noise $\sigma_{\alpha}$ is independent of $R_{\alpha}$, and is given by 
\begin{equation} \label{eq:noise}
    \sigma^2_{\alpha}=C/T_{\alpha}
\end{equation} 
for some coefficient $C$.  $T_{\alpha}$ is the computer time expended for configuration $R_{\alpha}$. 
This implies that the mean value of the matrix $A$, 
 \begin{equation}\label{eq:amat}
     \left<{\hat A}_{mn}\right>=
     M\left< \frac{f_m(R_{\alpha}) f_n(R_{\alpha})}{\sigma^2_{\alpha}} \right>_F
     =\frac{\sum_{\alpha} T_{\alpha} }{C} \int dR F(R) f_m(R) f_n(R) 
 \end{equation}
 where $\langle \dots \rangle_F$ means the average over the selection of points $R_{\alpha}$. We assume that $T_{\alpha}$ is constant or not correlated with $R_{\alpha}$.
Then it follows from Eq. (\ref{eq:amat}) that the mean value of ${\hat A}_{mn}$ depends only on the total computer time, $T=\sum_{\alpha} T_{\alpha}$. 
From Eq. (\ref{eq:llsq}), the error in the model will have the same property: the error in the fitted model is only proportional to $T^{-1/2}$: it does not depend on how many points we use to determine the data or how that time is allocated between the various configurations as long as the overall computer time is fixed. 
This conclusion holds when the configurations are independently sampled from some distribution, the model potential is determined by minimizing the mean squared deviation of the model energy from the noisy energy and the QMC errors are given by Eq. (\ref{eq:noise}).  

It is quite important to use forces in the training because they are much more plentiful than total energies by a factor of $3n_I$. Also forces give local information and integrate well with the ML models which are also local.  The likelihood function for optimizing the model including both energies and forces is
\begin{equation} \label{eq:loss}
{\cal O}= \chi^2_V + \lambda \chi^2_F.
\end{equation}
The likelihood function is the correct object to minimize if the noise is component-by-component independent and if the fluctuations are normally distributed. Both are reasonable assumptions for QMC data, though there can be deviations.
The parameter $\lambda$ should be set to one for maximum likelihood but it is often treated as a free parameter.
Including the contribution of the force data, the matrix determining the errors gets an extra contribution:
\begin{equation}
    {\hat A}_{mn} = \sum_{\alpha=1}^M \frac{f_m(R_{\alpha})f_n(R_{\alpha})}{\sigma_{V}^2}
    +\lambda\sum_{\alpha,\beta,i} \frac{\nabla_{\beta,i} f_m(R_{\alpha})\nabla_{\beta,i} f_n(R_{\alpha})}{\sigma_{F}^2}
\end{equation}
where $\nabla_{\beta,i}$ is the derivative of the $\beta^{th}$ Cartesian component on the $i^{th}$ atom.
The above discussion for the energy follows through.  The values of the matrix elements in $A$ have changed, becoming much larger, implying that the stochastic error on the model is much reduced, however the average matrix elements still only depend on the total computer time, not on how many configurations are present. 

In the above discussion, we have made some assumptions that need to be tested to see if they apply to realistic problems. What follows are two examples that show this property.  First, a model system of many-body silicon where we know the inter-atomic potential and, second, an actual QMC calculation of many-body hydrogen at high pressure. 

\section{Silicon example}

In this section, we give an example to illustrate the effect of stochastic error in training a neural network potential where the exact potential is known but where the potential has more realistic features.
We use the Tersoff potential~\cite{Tersoff1988} for Si, included in the LAMMPS program~\cite{LAMMPS}, as the ground truth. 
A MACE~\cite{MACE} model was trained on a set of configurations.
We do not expect that the MACE model will be able to capture the Tersoff potential perfectly though it should not be too difficult.

The training set was generated by isothermally compressing 64 Si atoms at $3000$ K from $0.467\rho_0$ to $1.282\rho_0$, where $\rho_0=4.994\times10^{-2}$~\AA$^3$ is the ambient density of Si solid. This training set provides a diverse set of local environments over a wide range of densities.
We used the MACE model~\cite{MACE} an equivariant message passing model that uses higher body order messages with $32$ invariant messages, angular resolution $l_{\text{max}} = 3$, and a cutoff radius of $3.2$~\AA, the same as the range of the Tersoff potential. Radial features are generated using $8$ Bessel basis functions and a 5$^{th}$ order polynomial envelope for the cutoff.
The loss function is a parameterized likelihood function with $\lambda=1.3$ \footnote{In the MACE input $\alpha_F=10$. To convert to the likelihood function $\lambda=\alpha_F(\sigma_F/\sigma_V)^2/(3n_I)$.}
Models were trained with AMSGrad variant of Adam, with default parameters, a learning rate of $0.01$ and a batch size of $4$.

We trained the model on different sizes of training sets ranging from 660 configurations to 4620 configurations.
We performed 3 different optimizations on the same data with those sizes. The first set used the clean data, data without noise, to get a baseline.
The trained model got better as more data was added, presumably by covering phase space better.  
Using the same configurations, we then added normally distributed noise to both the energies and forces but using a larger amount of noise as $M$ increased as detailed in Table (\ref{tab:tersoff}).
Because the statistical noise is inversely proportional to the square root of how long one samples in QMC, assuming a fixed total computation budget, the standard deviation of the noise on the energies and forces followed  $\sigma \propto \sqrt{M}$. For each configuration we subtracted the mean value of the force.
Without this modification, the potential energy would be inconsistent with the forces since any model that only contains differences of coordinates will have a zero mean force.

\begin{figure}[ht]
\includegraphics[width=0.95\linewidth]{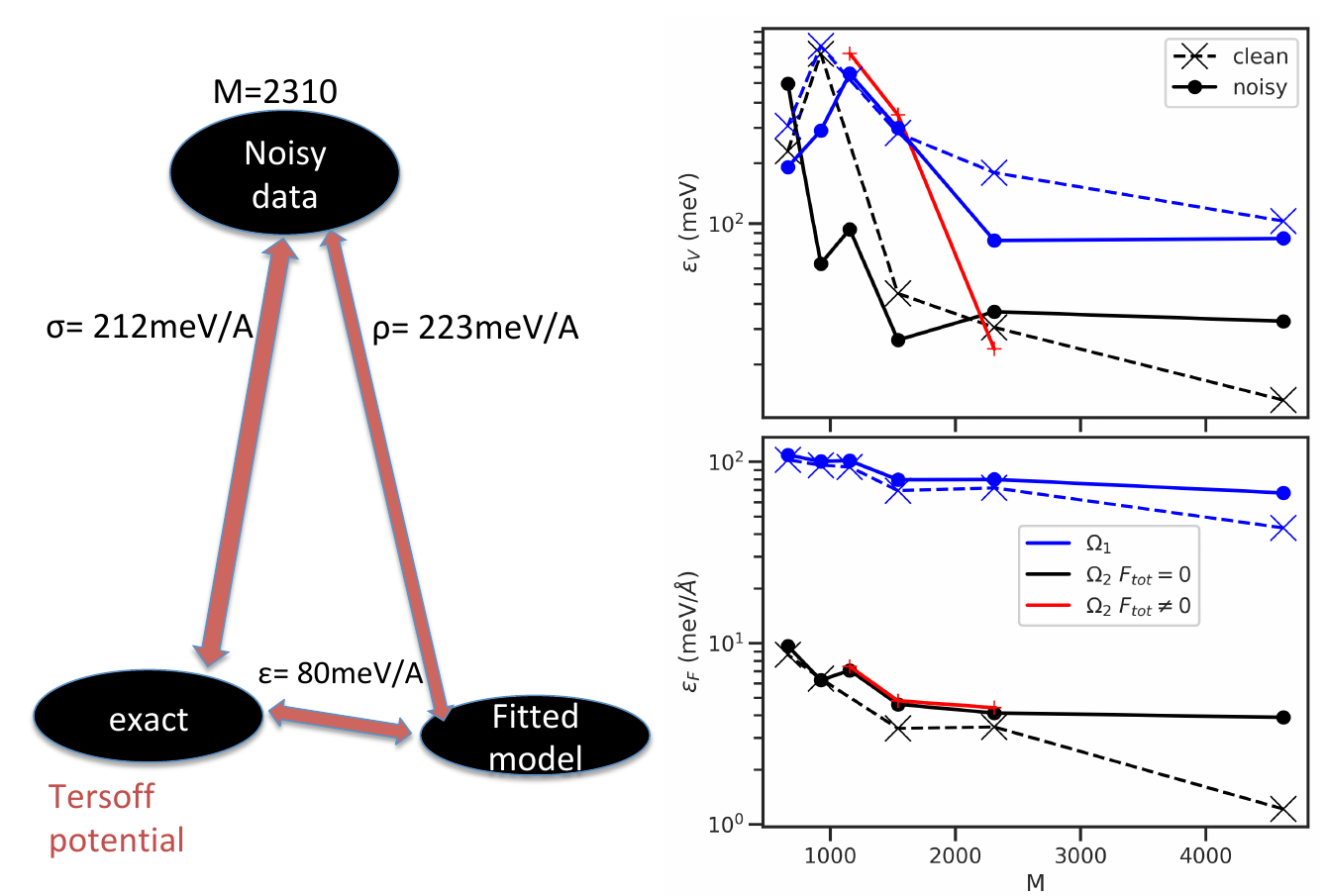}
\caption{On the left is an illustration of the force errors in the data and model. The numbers are for 2310 points in the training set evaluated on test set $\Omega_1$. The figure on the right shows how the errors depend on the number of sample points, whether noisy data or clean data was used for training and how the measured errors depend on the test set. The upper figure is the energy, lower figure the forces. Solid lines interpolate the noisy data, dashed lines the clean data. Black and red curves are evaluated on test set $\Omega_1$, blue curves on test set $\Omega_2$.
The red curve shows the model errors when we do not require the total force on the ions to be zero: the convergence of the energy is worse.}
\label{fig:tersoff}
\end{figure}

The values of $\sigma$, $M$ and the resulting $\epsilon$ are shown in Tables (\ref{tab:tersoff},\ref{tab:tersoff-train-rho0})  and in figure (\ref{fig:tersoff}).  Table (\ref{tab:tersoff}) gives errors as evaluated on an independent test set of 100 configurations (set $\Omega_1$).
This test set was generated with an MD trajectory obtained by heating a silicon crystal at ambient density from $1000$ to $3000$ K in $5$ picoseconds. This set was not used in the training.
Table (\ref{tab:tersoff-train-rho0}) shows the errors on a different test set (denoted as $\Omega_2$) which was a subset of 400 configuration of the training set with densities within $6\%$ of $\rho_0$. Strikingly different errors are obtained depending on the number of configuration used for the training and the makeup of the test sets. 

We see several things from these examples.  The most accurate model was obtained for the largest data set ($M=4620$) using the clean data for training, obtaining a model with $\epsilon_V^0=13$ meV and $\epsilon_F^0=1.2$ meV/\AA. Clearly, MACE is able to represent accurately the Tersoff potential near ambient conditions of density if given enough data. Note that the highest precision on the clean data was only achieved with relatively large data sets; the errors on the energy for smaller data sets were quite erratic. This could be a consequence of the value of $\lambda$ and the preponderance of force data.
On the same data with 60 meV noise, the errors of the model trained with noise were about three times as large.  The force errors showed a steady decrease as more data was added and accurate force models were obtained with fewer than one thousand configurations.  
This illustrates how much more information is provided by forces versus total energies and the insensitivity of the model to noise.

We also see that for the test set ($\Omega_1$) with densities close to ambient pressure of silicon, the force errors were 20 to 30 times smaller than for the subset of the training set ($\Omega_2$) that had a much greater variation in density. 
The model is much more accurate when tested over a very narrow range of densities even though it has been trained over a much wider range of densities. 

While the model energy was affected by the noise, except for the largest data set, the forces were not affected by the noise. 
Additional noisy data always improved the error in the forces even though the additional data had larger statistical errors. 
From the results of this example we conclude that one should always run with the maximum possible value of $M$ even though the resulting statistical errors are increased.  
Our analysis in the previous section of LLSF  remains valid for non-linear models. In practice, the best choice for $M$ will be influenced by the startup time of the QMC relative to the overall computer budget and the time to train the model. 

 We see from Table (\ref{tab:tersoff-train-rho0}), that the sensitivity parameter $\eta_F$ is small and decreases as $M$ increases. 
This implies that we can estimate the model error directly as $\epsilon_F = \lim_{M \rightarrow \infty} \sqrt{\rho_F^2-\sigma_F^2}$ based just on quantities that can be easily computed in a realistic model where we cannot compute the exact potential. However, the results for $\eta_V$, the energy value, are erratic and reinforce our finding that the energy model evaluated on test set $\Omega_2$ is not well converged.

In a third set of optimizations we did not subtract out the mean force. The results are  shown in Figure (\ref{fig:tersoff}), The force error was little affected but the energy error was increased when the training had fewer than 2000 configurations. 
It is possible that the choice of $\lambda$ in the loss function did not put enough importance on the energy errors versus the force errors.   

\begin{table}[h]
\centering
\begin{tabular}{|r|rrr|rrr|}
\hline
   M &  $\sigma_V$ & $\epsilon_V^0$&$\epsilon_V$ &  $\sigma_F$ &$\epsilon_F^0$ & $\epsilon_F$   \\
\hline
4620 &           60 &            13 &       33 &            300 &           1.2&           3.9  \\
2310 &           42 &            31 &       36 &            212 &           3.4 &          4.1  \\
1540 &           34 &            45 &       26 &            173 &           3.4 &          4.6  \\
1155 &           30 &          1214 &       94 &            150 &           6.2 &          7.1  \\
 924 &           26 &           697 &       63 &            134 &           6.4&           6.2  \\
 660 &           22 &           229 &      496 &            113 &           8.6&           9.6  \\
\hline
\end{tabular}
\caption{ Energy (in meV) and force errors (meV/\AA) as a function of the number of training configurations, $M$, for the Tersoff model. $\sigma_V$ is the standard deviation of the noise added to the energies and $\sigma_F$ to the forces. $\epsilon$ is the rms deviation of total energy and force components with respect to the exact values as evaluated on the test set described in the text. $\epsilon_V^0$ and $\epsilon_F^0$ are evaluated on the model trained with noiseless data.  Errors in this table were computed using the 100 configurations in the test set ($\Omega_1$,see text).  }
\label{tab:tersoff}
\end{table}

\begin{table}[h]
\centering
\begin{tabular}{|r|rrrrr|rrrrr|}
\hline
M & $\sigma_V$ & $\epsilon_V^0$ & $\epsilon_V$ & $\rho_V$ &  $\eta_V$ & $\sigma_F$ & $\epsilon_F^0$ & $\epsilon_F$ &  $\rho_F$ &  $\eta_F$ \\
\hline
4620 &      60(2) &  103 &      84(3) &   106(4) &     -0.06 &   297.3(8) &  43 &    67.3(4) &   301(1) &      0.013 \\
2310 &      43(2) &  179 &      82(4) &    95(4) &     -0.11 &   211.5(8) &  72 &    80.0(9) &   223(1) &      0.016 \\
1540 &      35(2) &  282 &     298(7) &   300(7) &      0.16 &   171.8(7) &  69 &      80(1) &   185(1) &      0.031 \\
1155 &      29(2) &  520 &    557(18) &  564(18) &     -4.55 &   148.4(7) &  94 &     101(2) &   177(1) &      0.025 \\
 924 &      29(2) &  763 &    290(13) &  286(12) &      1.83 &   133.7(9) &  96 &     100(3) &   163(2) &      0.040 \\
 660 &      24(2) &  306 &    191(11) &  193(12) &      0.05 &   112.5(7) &  103 &     109(2) &   152(2) &      0.060 \\
\hline
\end{tabular}
\caption{Errors on a subset of $400$ training configurations within 6\% of ambient density, test set $\Omega_2$. $\eta$ is a measure of the sensitivity of the energies and forces to the noise. $M$ is the number of configurations used in training.  Numbers in $(\ldots)$ are the statistical errors in the last digits. Other notation and units as in Table (\ref{tab:tersoff}).}
\label{tab:tersoff-train-rho0}
\end{table}

\section{Dense hydrogen example}

We now discuss a recent calculation~\cite{Niu2023} of many-body hydrogen under extreme conditions of temperature and pressure. Bulk hydrogen is important both because of its ubiquitous presence in the universe and in technological applications~\cite{RMP2012}. Even though it is the “holy grail” of high pressure studies, its phase diagram above 100 GPa is uncertain.  Experiments at high pressure are very difficult so they cannot be used to validate the computational approaches. 
Hydrogen is a good test case for computer simulation, QMC and machine learning because one does not need to use a pseudopotential so that does not complicate the algorithm or cause errors.
Although it does not have core electrons, its electronic structure has other complications such as a still elusive molecular-atomic transition that may coincide with its insulator-metal transition as well as effects caused by the very large quantum zero-point of the protons resulting in solid H$_2$ being disoriented at low temperatures.
QMC has been used extensively on this system~\cite{RMP2012}. Statistical errors are small with only a single electron per atom.
There is no fixed-node error for an isolated H$_2$ molecule if the 2 electrons have different spins. Within the molecular phase, wavefunction nodes are primarily in the low density region external to the molecules.  Hence we find the biases from the  fixed-node approximation is very small~\cite{Morales2014}. 

Fixed-node DMC (FN-DMC) calculations were done with QMCPACK version 3.9.2~\cite{Kim2018,Kent2020} using a Slater-Jastrow trial wavefunction with PBE orbitals computed using a Troullier-Martins PBE pseudopotential with core radius $r_c=0.37 a_{bohr}$  and a planewave cutoff of $200$ Ry.
The Jastrow functions contain two-body correlations as a sum of short-range and long-range contributions, both fully optimized for each configuration.
The FN-DMC calculations used $>2000$ walkers and were projected for $100$ ha$^{-1}$ with a timestep of $0.02$ ha$^{-1}$.
Canonical twist-averaging was performed using a $4\times4\times4$ shifted twist grid.
We used the Chiesa force estimator \cite{Chiesa2005} correcting the 
mixed-estimator bias with linear extrapolation.
Finite-size corrections were applied to the total energy~\cite{Chiesa2006,Holzmann2016} using the computed structure factor.
Forces are obtained from the twist-averaged electronic densities.  No further finite-size corrections were applied to the forces, as the leading order energy correction is a function only of the mean density  and thus does not affect the force. 

We generated a large set of hydrogen configurations, all containing 96 protons, using several different simulation methods: CEIMC, AIMD with DFT forces, both classical simulations and path integral simulations.  
The temperatures of the simulations ranged from 300K to 2200K, the pressures from 50GPa to 200GPa. The configurations included mostly molecular hydrogen, but for the higher temperatures and pressures configurations with dissociated molecules were present. 
The lower temperature simulations were of disoriented h.c.p. crystals. Above 1000K most configurations were of liquid H$_2$ or liquid H. 
The forces for 100,000 configurations were calculated using DFT (both PBE and vdW-DF1 functionals). We calculated QMC forces and energies for more than 16,000 configurations chosen over a range of temperatures and pressures. The data and information about the temperatures and pressures resides in a public data base~\cite{ythub}.

\begin{figure}[ht]
\includegraphics[width=0.52\linewidth]{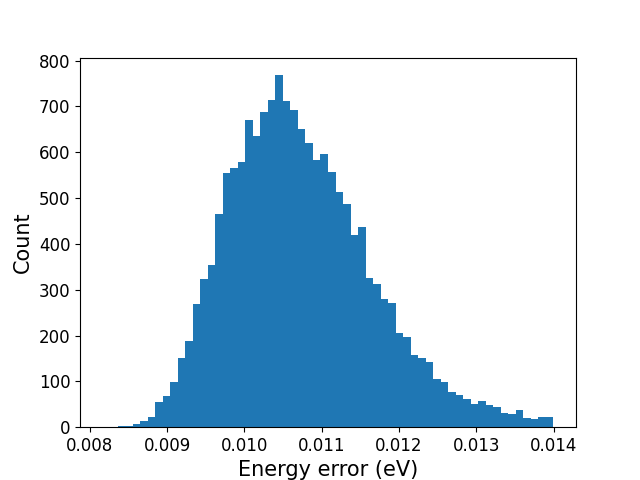}
\includegraphics[width=0.52\linewidth]{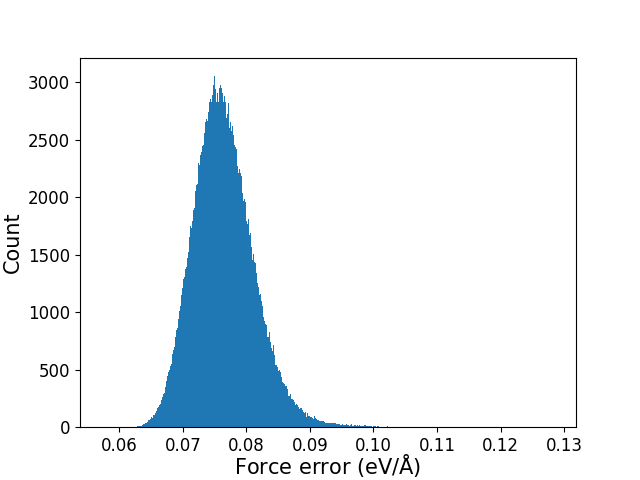}
\caption{Histograms of the estimated values of standard deviations of the total energy ($\sigma_V$) and force components ($\sigma_F$) from 16290 dense hydrogen configurations, each with 96 protons, within the pressure range 50GPa - 200GPa and temperature range 300K - 2200K. The standard deviations were estimated from the mean squared deviations of each individual DMC simulation.   We obtained $\sigma_v =$ 10.8 meV and $\sigma_F=$ 76.7 meV/\AA. A few percent of configurations with $\sigma_V> 0.014$ were discarded.}
\label{fig:Hydrogen_errors}
\end{figure}

Histograms of the potential and force errors are shown in Fig.~(\ref{fig:Hydrogen_errors}) as determined using a standard error analysis of the QMC calculations on each ionic configuration. Statistical analysis shows that the stochastic errors of the estimated forces and energies are close to normal justifying the use of the $\chi^2$ loss function. There were a few ($\approx 2\%$) configurations which had estimated energy errors twice the average which we discarded pending investigation into the cause of the larger errors. The variation of the remaining errors is less than 10\%.   
The error in the energy and force is only weakly dependent on the proton configuration within the range of pressures and temperatures studied and have mean standard errors of $\sigma_V=$10.8 meV on the total energy and $\sigma_F=$76.6 meV/\AA~ on each force component.
The mean value of the ionic forces within each configuration was subtracted from the individual ionic forces as was done in the silicon example. 

\begin{figure}[ht]
\includegraphics[width=.9\linewidth]{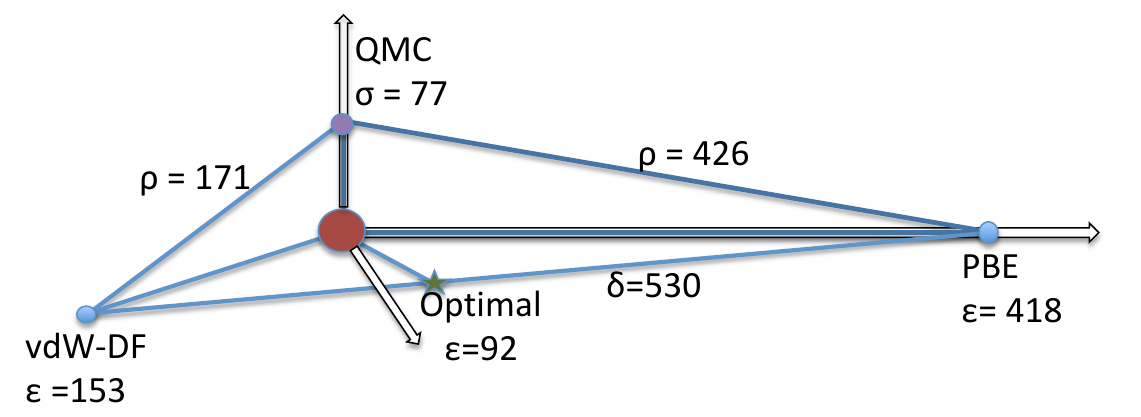}
\caption{Tetrahedron showing the QMC, PBE  and vdW rms force errors (in meV/\AA) for a subset of the hydrogen configurations for pressures 175GPa and 200 GPa and temperatures between 600 and 2000 K. The arrows represent an orthogonal set of axes. The red dot represents the ground truth, the purple dot the QMC error, the blue dots the PBE and vdW errors and the star, the optimal linear combination of PBE and vdW forces.   }
\label{fig:DFT}
\end{figure}

Using the QMC energies and forces we can make an estimate of the accuracy of DFT functionals and their error correlation.  We have previously benchmarked DFT functionals on dense hydrogen~\cite{Clay2014,Clay2016}. 
In that work, other methods were used to quantify the errors such as the mean absolute error.  Using the property that the QMC errors are uncorrelated with the DFT errors, we can find an estimate of $\epsilon$, the rms difference between DFT functionals and the ground truth; see table (\ref{tab:hydrogen errors}).  We agree with the conclusion in our previous study: for dense molecular hydrogen, the vdW-DF functional~\cite{vdw2004} has errors about one third of that of the PBE functional~\cite{PBE1996}. The rms  error of the total energy using the vdW functional is 1.24 eV. Because it is the variance of the energy fluctuations that is proportional to the size of the system, for a local energy error we divide by $\sqrt{n_I}=9.8$. This gives a local energy error of 0.127eV= 1440 K, reinforcing our conclusion that even vdW-DF is not accurate enough to predict the phase diagram of dense molecular hydrogen when energy scales of 1440K are relevant.  A diagram illustrating the magnitudes of the force errors in these 2 functionals is shown in Fig. (\ref{fig:DFT}).  It shows that the PBE force errors are anticorrelated with the vdW errors, though with three times their magnitude, agreeing with our experience that PBE is a better description of the metallic-atomic phase and vdW of the insulating-molecular phase.   By taking a linear combination of the two functionals (77\% vdW, 23\% PBE), one can get a significantly lower rms force error ($\epsilon_F$=92 meV/\AA) for this regime of temperature and pressure though with this combination the rms energy error is slightly higher.

\subsection{ DeePMD model} Our first ML model for dense hydrogen used a  $\Delta$-learning model containing three terms:
\begin{equation} \label{eq:final-model}
E = E_{\text{pair}} + \Delta E_{DFT} + \Delta E_{DMC}
\end{equation}
where $E_{\text{pair}}$ is the contribution from an analytic potential using only the repulsive part of the exact BO energy of an isolated H$_2$ molecule~\cite{Kolos1964}. 
This allows an accurate description of the internal motion of a single H$_2$
without having to learn it from the training data.
The next two terms were determined by a ML model.
$\Delta E_{DFT}$  is a model trained on DFT energies, forces, and stresses (either PBE or vdW functionals) minus the pair potential contribution and was trained on roughly 100,000 hydrogen configurations. 
This model should capture the overall band structure and electron correlation effects as described by the density functional.
Finally, $\Delta E_{DMC}$ is determined by fitting $E - E_{\text{pair}} - \Delta E_{DFT}$ to the subset of 20,000 configurations for which we have computed QMC forces.  We expect $\Delta E_{DMC}$ to be a small correction: a smooth function of proton positions that can be described by a simpler ML model with fewer parameters, 
thus making good use of the computationally expensive and noisy QMC forces.

To model $\Delta E_{DFT}$ and $\Delta E_{DMC}$ we used the Deep Potential Molecular Dynamics (DeePMD)~\cite{Han2018,Zhang2018a,Zhang2018} version 1.3.3. 
Symmetry invariant feature vectors for atomic environments are constructed via the ``smooth edition''~\cite{Zhang2018} descriptors with angular and radial information.
The angular features are smoothly cut-off from $3.4$ to $4.0$~\AA, and radial features from $5.1$ to $6.0$~\AA.
We used $8$ neurons in the embedding layer, and $3$ layers of $8$ hidden neurons each for the descriptor network.
The DeepMD training consisted of $10^6$ iterations using 8 configurations to update the neural network parameters at each iteration. The starting and ending learning rates were set to $5\times 10^{-3}$ and $2\times 10^{-7}$. 

\subsection{ MACE model}

Our second optimization used the MACE model~\cite{MACE}, an equivariant message passing model that uses higher-body order messages.   We did not use $\Delta$-learning model of Eq. (\ref{eq:final-model}) with MACE. Instead the QMC data were fit directly without a baseline or DFT layer.
Optimization used $64$ invariant messages, an angular resolution $l_{\text{max}} = 3$, and a cutoff radius of $5$~\AA. 
The loss function is the parameterized likelihood function with forces weighted with $\lambda=3.5$.

\subsection{ Errors in the DeePMD and MACE models}

\begin{figure}[ht]
\includegraphics[width=0.9\linewidth]{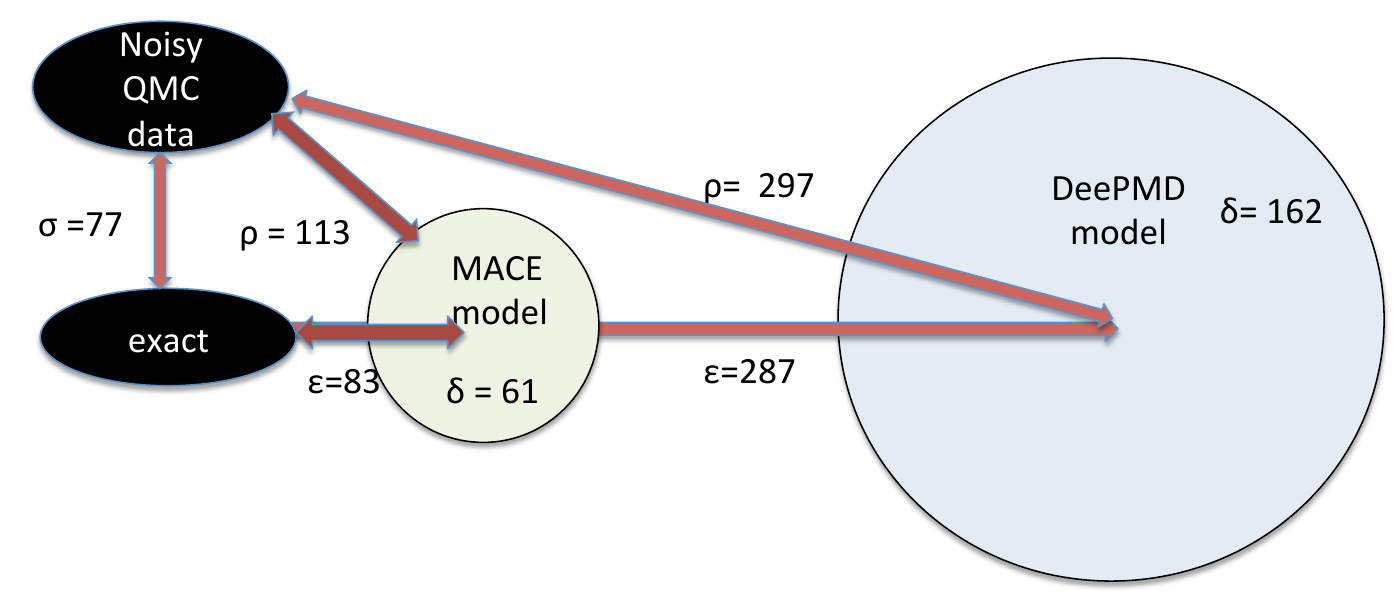}
\caption{  Diagram illustrating the magnitude of r.m.s. errors in forces (in meV/\AA) for the QMC forces and 2 ML models for dense hydrogen, a 3-level DeepMD trained on DFT forces and on QMC forces and a single level MACE model trained only on QMC forces.  Circles indicate the magnitudes of the cross-validation uncertainty in the DeepMD and MACE models.}
\label{fig:Hydrogen}
\end{figure}

Results in Table (\ref{tab:hydrogen errors}) give the fitting error $\rho$ (Eq.~(\ref{eq:rho_V})) and cross-validation error $\delta$ (Eq.~(\ref{eq:delta_V})) in the energies and forces for the two models used and for the clean DFT calculation.  A visual depiction is shown in Fig. (\ref{fig:Hydrogen}).
We see for both models that $\rho > \sigma $.  This means that the models have not exhausted the information in the QMC data, i.e. the values of $\chi^2$ are large compared to the number of data points.  In particular, the QMC energy uncertainty of 10meV is two orders of magnitude smaller than the fitting error of the models to the QMC energies.  (We remind the reader that what is quoted for $\sigma_V$ is the standard deviation on the total energy of 96 protons in the supercell.)
The energies are much worse in this regard than the forces.   The MACE model force errors are only 46$\%$ larger than the force QMC errors, so the force information from the QMC data has been reasonably well fit by the models. This indicates that more research is needed to find a more accurate model for dense hydrogen in this regime of temperature and pressure.  We estimated the accuracy of the models with respect to the exact energy and forces using Eq. (\ref{eq:rhoepsneta}) assuming $\eta=0$. 

The different performance of the energy and force errors is not understood. One might think that as a model approaches the exact potential, that the force and energy errors would decrease together. However, for the two DFT functionals $\epsilon_V/\epsilon_F  = 8$\AA ~but for the 3 DeePMD models this ratio is about 4.5\AA ~while for the MACE model it is 13\AA. 
Even though the forces drive the trajectory in MD, it is the potential that determines the equilibrium properties, for example the free energy.

There is a large uncertainty, $\delta$, in the ML models as determined by the 5-fold cross validation also shown in Fig. (\ref{fig:Hydrogen}). The uncertainty of the MACE model is less than half that of the DeePMD model, but it is of the same magnitude as the errors in the model forces. This indicates that the training would profit from more data; the existing ML models are still highly sensitive to the QMC noise and the selection of configurations. 

The QMC/MACE model has the smallest errors in both the energy and forces from the other models including the clean vdW DFT model. However, this particular trained model has not yet been tested with MD or PIMD simulations. It is not unusual for a MD or PIMD simulation to reveal hidden problems with an energy surface when a trajectory ventures into regions of configuration space that the model has not been trained for.   It is clearly worthwhile to do such tests since the MACE scales better with the number of protons and is several orders of magnitude faster than using DFT functionals or Coupled Electron Ion MC (CEIMC) for studies of dense hydrogen. 

\begin{table}[]
    \centering
    \begin{tabular}{|c|c|r r r|r r r|}
    \hline
       1& 2& $\rho_V$& $\epsilon_V$ & $\delta_V$ &$\rho_F$& $\epsilon_F$ & $\delta_F$ \\
         \hline
       PBE &&  3360 & 3360& & 426 & 418& \\
       vdW && 1240 & 1240 & & 171 & 153 & \\\hline
       PBE/DeePMD&  & 1388  & 1388& & 306 & 296 & \\
       vdW/DeePMD & & 1412   & 1412 & & 329 & 320& \\\hline
       vdW/DeePMD&QMC/DeePMD    & 1268   & 1268 & 312& 297 &287 & 162\\
       QMC/MACE && 1109  &1109 & 226& 113&83& 61 \\
        \hline
    \end{tabular}
    \caption{RMS errors in total energies (in meV) and forces (in meV/\AA) with respect to QMC calculations using 16,290 configurations of 96 protons at pressures between 50GPa and 200 GPa. The first two columns abbreviate the construction of the models. For example ``PBE'' means that the errors wrt QMC energies are compared to PBE energies, ``PBE/DeePMD'' means that a DeePMD model is constructed using PBE forces and compared to QMC, and ``QMC/DeePMD'' means a second level DeePMD model trained using QMC energies is added to the first level model. Symbols ($\rho,\epsilon,\delta$) are defined in Fig. (\ref{fig:Diagram1}). To compute $\epsilon$ we assumed $\eta=0$ and used $\sigma$ from Fig. (\ref{fig:Hydrogen_errors}). $\delta$ was estimated using a 5-fold cross-validation (jackknife) procedure. }
    \label{tab:hydrogen errors}
\end{table}

\section{Summary and Future Research}
We have shown that it is feasible to use QMC forces to train state-of-the-art ML models. Such models trained on QMC energies and forces open the
possibility to quantify the accuracy of  models used in Monte Carlo or Molecular Dynamic studies.  We have shown in the molecular hydrogen example, that the DeePMD and MACE models have not fully exploited the information contained in the QMC force and energy data.

Our analysis of the errors in the two analytic examples give some guidance on how to use QMC forces to improve ML models. 
Clearly one needs to calculate forces, not just energies because QMC forces have much more information to be exploited.  
It is better to have forces on many configurations even though the force errors are larger than if fewer configurations were examined.  
A model is well converged if all of the important parameters are well trained. This can be judged by a cross-validation analysis.  
Models with fewer parameters can become well-converged more easily. The ratios, $\rho_V/\sigma_V$ and $\rho_F/\sigma_F$, can be used to judge whether models have extracted all the information from the data.  These are some of the ways noisy data can help in validating machine learning models.
They present great opportunities but there are many open research problems. 

It is not clear which type of QMC will be most useful to generate data for machine learning: VMC, DMC, or RQMC?   
VMC can generate uncorrelated forces more efficiently. However, unless extensive careful optimization is done for each configuration, it can have a much larger systematic bias.   
DMC and RQMC are automatic procedures for eliminating much of this bias and should be equally accurate for the energy. A systematic examination of their efficiency and bias for forces has not been reported. We can also ask whether PIMC will be useful in the warm dense matter regime.

There has been limited research on the efficiencies and biases of the suggested estimators for forces such as the  Chiesa~\cite{Chiesa2005} estimator or zero-variance, zero-bias estimators~\cite{Assaraf2000}. 
Is extrapolation of the forces with DMC adequate or do we need to do forward walking, reptation  or use unbiased estimators? What is systematic bias of the forces coming from estimators and from the fixed-node or fixed-phase approximations?
Can second derivatives of the energy computed by QMC be useful to train the models?  At the moment we do not know their statistical errors and biases.

An important practical questions is how to speed up the QMC calculation so as to deliver data more efficiently.  We found in the Silicon example that it is best to have as many points as possible even though the noise would be larger. 
If we minimize the start-up cost of a run, we can generate more data but we need robust but ``lightweight'' QMC methods for this.
It would be best to avoid a lengthy optimization of the trial function.  
Typically, QMC uses DFT orbitals for the trial wavefunction, however there are drawbacks to tying the QMC run to a fully converged DFT calculation as that would decrease its efficiency. One possibility is to develop a simplified single-particle theory, for example by using optimized effective interactions~\cite{Pierleoni2008}.
One gains with a compact expansion of the electron orbitals in terms of a basis, first, because determining the best expansion coefficients is rapid, and, second, the evaluation of the trial wavefunction and its derivatives within the QMC calculation is faster. 
Such a description is possible because the electron-ion Jastrow factor and backflow already contain the wavefunction cusps and the dynamic correlations of the electrons and do not need to be present in the orbitals. 
What is important is to find orbitals with a good representation of the nodal surfaces so that the fixed-node approximation is accurate. 
It has been found that analytic Jastrow functions, backflow functions and three-body functions can be used~\cite{Holzmann2003} and need little or no optimization.  Alternative ways to antisymmetrize the trial function such using geminal orbitals or pfaffians might prove useful. In CEIMC calculations of dense hydrogen, we made minor tweaks to the analytic functions for a given density and did not reoptimize for small changes in density or temperature~\cite{Pierleoni2008}.

What is the trade-off between high quality data and covering configuration space better?  In the Tersoff Silicon example discussed above, we found that the best models come from the largest data set, even though the noise in forces was larger.
What is the best way to select configurations?  Can we formulate in terms of a sampling function F(R) or do learning procedures improve on this?  For example, should one keep the configurations spaced out in configuration space.

What type of model should we to fit to? We have tried fitting directly to the QMC data and to fitting to a  3 level $\Delta$-model containing molecular forces, DFT forces and QMC forces. It seems useful to start with a simple model that prevents atoms from getting too close to each other.  But the MACE example for molecular hydrogen shows that hierarchical model may not be necessary.  Having a model that needs accurate DFT calculated forces will make the data generation less efficient. Perhaps an embedded atom or tight-binding model will be a useful alternative baseline model. There is also an advantage to a model with fewer free parameters since having more parameters requires more QMC data for convergence. 

Incorporating known physical constraints on the models can be important.  We have already used short-ranged interactions such as the molecular H$_2$ binding curve in our model.  
However, long-ranged interactions are essential for calculations of the electrostatic properties such as the dielectric constant. But since forces on atoms are dominated by short-ranged interactions and ML models typically have a cutoff at a small multiple of the interatomic distance, long-range interactions may not be correctly determined in the ML model.
In many cases, one is able to calculate parameters of the long-range interactions from first principles. 
For example, the dispersion interaction only depends on the multipole moments of the atoms or molecules present and their polarizabilities, and can be determined using QMC methods on small supercells.  
On the other hand in metals, the bare Coulomb interaction is screened by the mobile charges; models to describe the magnitude of the long-range interaction are more complex. 
A challenging problem is to simulate a metal-insulator transition such as occurs when hydrogen changes from the molecular phase to the atomic phase.    
The character of the long-range interaction changes during the transition and is correlated with the changes in the short-ranged structure of the system. 
It is important to be able to model the interactions at all distance scales accurately and self-consistently in order to make precise predictions of this transition in the ML model.

We have demonstrated that QMC is useful in generating data for machine learning potentials and assessing the error of those potentials. In principle, the data should have a much smaller systematic bias.  The noisy feature of QMC forces and energies can be an advantage both in allowing more complete sampling of phase space, in optimization of the model and assessing the model errors. There is much scope for improving the methodology to make QMC a useful tool for generating ML data. 

\section{Acknowledgements}
We acknowledge useful discussions with L. K. Wagner, E. Ertekin, and  K. Ly.
SJ and DC were supported by DOE DE-SC0020177. CP was supported by the European Union - NextGenerationEU under the Italian Ministry of University and Research (MUR) National Innovation Ecosystem grant ECS00000041 - VITALITY - CUP E13C22001060006. The Flatiron Institute is a division of the Simons Foundation. Computations were done on Blue Waters Computer and the Illinois Campus Cluster, supported by the National Science Foundation (awards OCI-0725070 and ACI-1238993), the
state of Illinois, the University of Illinois at Urbana-Champaign and its National Center for Supercomputing Applications  and at the Oak Ridge Leadership Computing Facility, supported under Contract DE-AC05-00OR22725, and the Flatiron Institute Rusty cluster. 

\section{References}
\bibliography{ref}

\end{document}